\pdfoutput=1
\documentclass[11pt,a4paper]{article}

\usepackage{jheppub}
\usepackage{microtype}
\usepackage[utf8x]{inputenc}
\usepackage{booktabs}
\usepackage{multirow}

\graphicspath{{figures/}}

\makeatletter
\g@addto@macro\bfseries{\boldmath}
\makeatother

\DeclareMathOperator{\im}{Im}
\DeclareMathOperator{\re}{Re}
\DeclareMathOperator{\sgn}{sgn}

\DeclareMathOperator{\tr}{Tr}
\DeclareMathOperator{\Li}{Li}
\newcommand{\MS}{$\overline{\text{MS}}$}

\title{Life of $\Pi$}
\preprint{\mbox{}\hfill DESY 17-154\\\mbox{}\hfill IPPP/17/71}

\author[a]{Andreas Maier,}
\author[b]{Peter Marquard}

\affiliation[a]{Institute for Particle Physics Phenomenology, Durham University, Durham, United Kingdom}
\affiliation[b]{Deutsches Elektronen Synchrotron (DESY), Platanenallee 6, Zeuthen, Germany}

\emailAdd{andreas.maier@durham.ac.uk}
\emailAdd{peter.marquard@desy.de}

\keywords{Perturbative QCD, Heavy Quarks}

\abstract{
  The heavy-quark contribution to the polarisation function $\Pi$ at
  higher perturbative orders is presently only known approximately. We
  scrutinise the accuracy of state-of-the-art approximations
  at three- and four-loop order. At three loops, we present for the first
  time a result with arbitrary numerical precision for general kinematics
  and compare to the best Pad\'e estimate. At four loops, we calculate the
  fourth (inverse) moment of the non-singlet heavy-quark vacuum
  polarisation in order to test the prediction for this moment based on
  Pad\'e approximation.
}

\begin{document}
\maketitle

\section{Introduction}
\label{sec:intro}

Vacuum polarisation is one of the earliest and phenomenologically most
important predictions of quantum electrodynamics (QED). Consequently,
the computation of the two-loop perturbative correction to this effect
constitutes one of the very first multi-loop calculations performed
within QED~\cite{Kallen:1955fb}.

Quantum corrections mediated through virtual quarks are of special
interest. They are closely connected to the total inclusive
hadron production cross section at lepton colliders through a dispersion
relation~\cite{Novikov:1977dq}. Conversely, it follows from the optical
theorem that, up to a simple normalisation factor, the cross section is
equal to the imaginary part of the quark contribution to the vacuum
polarisation. More precisely, the heavy-quark polarisation function
$\Pi$ and the cross section are related via
\begin{equation}
  \label{eq:disp_rel}
  \Pi(s') = \frac{s'}{12\pi^2}\int_0^\infty
  \text{d}s\,\frac{R(s)}{s(s-s')}\,,\qquad
  R(s) = 12 \pi \im \big[\Pi(s+i\epsilon)\big]\,,
\end{equation}
where the R-ratio for a heavy quark $Q$ is defined as $R(s) = \sigma(e^+
e^- \to Q\bar{Q}X)/\sigma_0$ with $\sigma_0 =
\tfrac{4\pi\alpha^2}{3s}$. Starting at four loops in the perturbative
expansion of the polarisation function, there is a contribution from
flavour-singlet diagrams with massless cuts.\footnote{The three-loop
  singlet diagrams vanish identically. This follows from Furry's theorem.} These cuts do
not correspond to the production of heavy quarks. In the following, we
will therefore restrict ourselves to the discussion of the
non-singlet polarisation function.

In the limit where the center-of-mass energy is far above both the scale
of non-perturbative dynamics and the masses of the quarks, the
polarisation function is known at four-loop
order~\cite{Baikov:2012zm}. The closely connected Adler function $D(s) =
-12\pi^2 s\tfrac{\text{d}}{\text{d}s}\Pi(s)$ is even known at five-loop order
for massless quarks~\cite{Baikov:2010je,Baikov:2012zn}. The
dimensionless polarisation function can only depend on the energy through
logarithms, which in turn give rise to the complete
imaginary part of the polarisation function. Thus, as per the optical
theorem Eq.~(\ref{eq:disp_rel}), the knowledge of the five-loop Adler
function allows a N$^4$LO prediction of the total quark production cross section.

However, in the production of heavy quarks the approximation of small
quark masses is not always justified. In fact, sufficiently close to the
production threshold the full quark mass dependence has to be taken into
account. A prominent scenario is the production of top--antitop pairs at
the projected first stage of CLIC at a center-of-mass energy of
$380\,$GeV~\cite{CLIC:2016zwp}. For the determination of the charm- and the bottom-quark
mass it is even the opposite limit of large quark masses (or small
center-of-mass energies) that is most
relevant. The coefficients in such a low-energy expansion can be
identified with (inverse) moments of the heavy-quark production cross
section via the aforementioned dispersion relation. These moments in
turn are the main ingredient in sum-rule determinations of the quark
masses~\cite{Novikov:1976tn,Novikov:1977dq}.

In the kinematic region where the quark mass is non-negligible much less
is known about the vacuum polarisation corrections than in the limit of
massless quarks. The first major step towards obtaining the three-loop
corrections was taken about 20 years ago~\cite{Chetyrkin:1995ii}, when
expansions in the low-energy, threshold, and high-energy kinematic
regions were exploited to construct a Padé-based approximation. Since
then, many more terms in the low-energy and high-energy expansions have
become
available~\cite{Chetyrkin:1996cf,Boughezal:2006uu,Maier:2007yn,Maier:2011jd},
allowing a systematic improvement of the approximation (see
e.g.~\cite{Masjuan:2009wy}). An alternative approximation procedure
based on Mellin-Barnes transforms was explored
in~\cite{Greynat:2010kx}. Independently, the cross section corresponding
to the imaginary part of the vacuum polarisation was computed
numerically in~\cite{Gao:2014nva,Chen:2016zbz}. Corrections involving
both massive and massless quarks were obtained already much earlier
in~\cite{Hoang:1995ex}.

At four-loop order, the same approaches were again used for an
approximate reconstruction of the heavy-quark corrections to the vacuum
polarisation~\cite{Hoang:2008qy,Kiyo:2009gb,Greynat:2010kx,Greynat:2011zp}. These
approximations were in turn expanded again in the low-energy limit in
order to obtain estimates for higher moments used in sum-rule
analyses. In the most precise determinations of the charm- and
bottom-quark masses from relativistic sum rules to
date~\cite{Chetyrkin:2009fv,Chetyrkin:2010ic,Dehnadi:2011gc,Bodenstein:2011ma,Bodenstein:2011fv,Narison:2011xe,Dehnadi:2015fra,Erler:2016atg}
the exactly computed first three physical
moments~\cite{Chetyrkin:2006xg,Boughezal:2006px,Maier:2008he,Maier:2009fz}
were considered together with an estimate of the fourth
moment.

To summarise, current knowledge of quark-mass corrections to the vacuum
polarisation at three- and four-loop order is based to a large degree on
approximations. If and in which sense approximations based on the scheme
considered in~\cite{Chetyrkin:1995ii,Hoang:2008qy,Kiyo:2009gb} converge
to the true results as more information is added is an open question,
which we do not intend to address in this work. Our goal is rather to
analyse to which extent approximations based on current knowledge and
their heuristic error estimates can be relied on. We aim to ameliorate
the dependence on approximations by providing new exact results at three
and four loops. At three-loop order we numerically calculate the vacuum
polarisation for general kinematics and compare to a new Padé-based
approximation constructed from many coefficients in the low- and
high-energy expansions as well as to the approximation obtained
in~\cite{Greynat:2010kx}. At four loops, we present an analytic result
for the fourth term in the low-energy expansion and compare to the
various estimates based on the
approximations~\cite{Hoang:2008qy,Kiyo:2009gb,Greynat:2011zp} to the
four-loop polarisation function.

\section{Conventions}
\label{sec:conventions}

The quark contribution to the vacuum polarisation is given by the
correlator of two vector currents, viz.
\begin{equation}
  \label{eq:Pidef}
  \Pi_{\mu\nu}(q) = (-q^2 g_{\mu\nu} + q_\mu q_\nu)\Pi(q^2) =  i \int \text{d}x\, e^{iqx} \langle 0 |T j_\mu(x)
                    j_\nu(0)|0\rangle\,,
\end{equation}
where the vector current is $j_\mu = \bar{\psi} \gamma_\mu \psi$. The
polarisation function $\Pi$ is conventionally renormalised in the
on-shell scheme, so that $\Pi(0) \equiv 0$. Its perturbative expansion
in the strong coupling constant $\alpha_s$ for a quark with charge $e_Q$
can be written as
\begin{equation}
  \label{eq:Pi_pert}
  \Pi(q^2) = \frac{3 e_Q^2}{16\pi^2}\sum_{i=0}^\infty \Pi^{(i)}(q^2)
  \bigg( \frac{\alpha_s}{\pi}\bigg)^i\,.
\end{equation}
We set the renormalisation scale $\mu=m_Q$, where $m_Q$ is the quark
mass renormalised in the on-shell scheme~\cite{Tarrach:1980up,Gray:1990yh,Chetyrkin:1999ys,Chetyrkin:1999qi,Melnikov:2000qh,Marquard:2007uj,Marquard:2015qpa,Marquard:2016dcn}.

In the following we are interested in the three-loop coefficient
$\Pi^{(2)}$ and the four-loop coefficient $\Pi^{(3)}$. The kinematic
dependence of the polarisation function is described by a single ratio
of energy and mass, which we define as $z = q^2/(4m_Q^2)$. We consider
the general case of complex $z$, which is needed for example when
describing unstable quarks, like the top quark. The perturbative
coefficients of the polarisation function are analytic functions of $z$,
apart from a branch cut along the positive real axis. Since we neglect
contributions from diagrams with massless cuts, the branch cut starts at
the open quark production threshold $z = 1$. The low-energy expansions
\begin{equation}
  \label{eq:Pi_le} \Pi^{(i)}(q^2) = \sum_{n=1}^\infty C_n^{(i)} z^n
\end{equation}
of the perturbative coefficients therefore converge for $|z| < 1$. In close
analogy, we write the expansions in the threshold region $z \to 1$ and
the high-energy region $z \to - \infty$ as
\begin{align}
  \label{eq:Pi_thr}
  \Pi^{(i)}(q^2) ={}& \sum_{n=1-i}^\infty\sum_{m\ge 0} K_{n,m}^{(i)}(1-z)^{\frac{n}{2}}\log^m(1-z)\,,\\
  \label{eq:Pi_he}
  \Pi^{(i)}(q^2) ={}& \sum_{n=0}^\infty\sum_{m\ge 0} D_{n,m}^{(i)} z^{-n}\log^m(-4z)\,.
\end{align}
The three-loop coefficients $C_n^{(2)}, D_{n,m}^{(2)}$ are known up to
$n=30$~\cite{Boughezal:2006uu,Maier:2007yn,Maier:2011jd}. At four loops,
the coefficients $C_n^{(3)}$ have been computed for $n =
1,2,3$~\cite{Chetyrkin:2006xg,Boughezal:2006px,Maier:2008he,Maier:2009fz}. The
threshold coefficients $K_{n,m}^{(i)}$ can be extracted from
calculations in a non-relativistic effective
theory~\cite{Hoang:2000yr,Pineda:2006ri,Beneke:2015kwa}; explicit
expressions obtained from NNLO results are given
in~\cite{Hoang:2008qy,Kiyo:2009gb}.

\section{Calculational setup}
\label{sec:setup}

We generate the diagrams contributing to the polarisation function with
\texttt{QGRAF}~\cite{Nogueira:1993}, obtaining 36 diagrams at three
loops and 700 diagrams at four loops. For inserting the Feynman rules,
evaluating traces, and performing general symbolic manipulations we use
\texttt{FORM}~\cite{Vermaseren:2000nd}. Colour factors are computed with
the \texttt{color}~\cite{vanRitbergen:1998pn} package. At four loops, we also perform an expansion
around $z=0$ up to order $z^4$. The resulting scalar integrals are
reduced to master integrals by exploiting integration-by-parts
identities~\cite{Chetyrkin:1981qh} according to
Laporta's algorithm~\cite{Laporta:2001dd} as implemented in
\texttt{Crusher}\footnote{\texttt{Crusher} uses
  \texttt{fermat}~\cite{fermat} and \texttt{GiNaC}~\cite{Bauer:2000cp}.}~\cite{crusher}.

At four loops, the expansion around $z=0$ results in vacuum integrals,
and the resulting master integrals are known
analytically~\cite{Chetyrkin:2006dh,Schroder:2005va,Schroder:2005hy,Laporta:2002pg,Chetyrkin:2004fq,Kniehl:2005yc,Schroder:2005db,Bejdakic:2006vg,Kniehl:2006bf,Kniehl:2006bg}. At
three loops, we derive differential
equations~\cite{Kotikov:1990kg,Remiddi:1997ny,Gehrmann:1999as} for the
master integrals expanded in the dimensional regularisation parameter
$\epsilon$. We solve the differential equations using the
Runge-Kutta-Dormand-Prince~\cite{dopri} method as implemented in the \texttt{odeint}
C++ library~\cite{odeint}. As boundary condition we choose values of the integrals at
$z_0 \approx 0$, which we obtain from the low-energy expansion performed
in~\cite{Maier:2007yn}. Note that we avoid $z_0 = 0$, since the
differential equations exhibit a singularity at this point. For general
complex $z$, we integrate the differential equations along a straight
line from $z_0$ to $z$. However, there are further singularities along
the positive real axis, even below the physical branch cut starting at
$z=1$. When $z$ is close to the real axis, we therefore perform a
contour deformation into the complex plane. In principle, any path that
bypasses the singularities is sufficient. In practice, we choose a
piecewise linear path from $z_0$ over $\re(z_0) + i \sgn(\im(z)) \re(z)$
and $\re(z) + i \sgn(\im(z)) \re(z)$ to $z$.

\section{Three-loop quark contribution to the polarisation function}
\label{sec:results}

In the following, we present our new result for the three-loop
polarisation function and compare to approximations based on previously
known expansion coefficients.

\subsection{Comparison to Padé-based approximation}
\label{sec:pade_comp}

We construct Padé-based approximants according to the
procedure described in~\cite{Baikov:2013ula}. We briefly summarise the
main aspects. First, we use subtraction functions listed
in~\cite{Baikov:2013ula} to split $\Pi^{(2)}$ into two parts,
\begin{equation}
  \label{eq:pi2_split}
  \Pi^{(2)} = \Pi^{(2)}_{\text{reg}} + \Pi^{(2)}_{\text{log}}\,,
\end{equation}
where all known logarithms and poles in the threshold and
high-energy expansions Eqs.~(\ref{eq:Pi_thr}), (\ref{eq:Pi_he}) have
been absorbed into $\Pi^{(2)}_{\text{log}}$. We then make a Padé ansatz
of the form
\begin{equation}
  \label{eq:Pade}
  [n/m] = \frac{\omega^n + \sum_{i=0}^{n-1} a_i \omega^i}{\sum_{i=0}^m b_i \omega^i}\,,
\end{equation}
where the variable $\omega$ is defined by the relation
\begin{equation}
  \label{eq:w_def}
  z(\omega) = \frac{4\omega}{(1+\omega)^2}\,.
\end{equation}
The approximants $[N/0], [N-1/0]$ are fixed by requiring that the
coefficients match the terms in the Maclaurin series of
\begin{align}
  \label{eq:pade_fix}
  P_{30}(\omega) ={}&
  z(\omega)^{31}\bigg(\Pi^{(2)}_{\text{reg}}\big(z(w)\big)-\sum_{i=0}^{30}
  \frac{H_i^{(2)}}{z(\omega)^i}\bigg)\,,\notag\\
  H_i^{(2)} ={}&
  \frac{1}{i!}\bigg(\frac{\partial}{\partial(1/z)}\bigg)^i\Pi^{(2)}_{\text{reg}}(z)\bigg|_{z\to \infty}\,.
\end{align}
The degree $N$ corresponds to the number of known coefficients
$C^{(2)}_n, D^{(2)}_{n,0}$, so $N=61$. Note that the threshold expansion
Eq.~(\ref{eq:Pi_thr}) is only used in the construction of
$\Pi^{(2)}_{\text{log}}$. In particular, terms that are analytic in
$\sqrt{1-z}$ are not considered for the approximation.

Further approximants are then constructed with the recurrence relations~\cite{baker:1970}
\begin{align}
  \label{eq:pade_recursion}
[N-\tfrac{j}{2}/\tfrac{j}{2}]
&= \frac{\hat{\eta}_j}{\hat\theta_j}
= \frac{\hat\eta_{j-2} -
  \omega\*\,\hat\eta_{j-1}}{\hat\theta_{j-2} -
  \omega\*\,\hat\theta_{j-1}} & j \text{ even}\,,\\
[N-\tfrac{j+1}{2},\tfrac{j-1}{2}]
&= \frac{\hat{\eta}_j}{\hat\theta_j}
= \frac{\hat\eta_{j-2} -
  \hat\eta_{j-1}}{\hat\theta_{j-2} -
  \hat\theta_{j-1}} & j \text{ odd}\,,
\end{align}
where $\hat{\eta}_j$ is the numerator of the approximant in the form of
Eq.~(\ref{eq:Pade}) and $\hat{\theta}_j$ its denominator. We discard all
approximants with poles inside the unit circle, which translate to
unphysical poles in the variable $z$.

Instead of constructing new approximants for various fixed numbers
$n_l$ of massless quark flavours, we decompose
\begin{equation}
  \label{eq:Pi_nl}
  \Pi^{(2)} = \Pi^{(2)}_{n_l^0} + n_l\,\Pi^{(2)}_{n_l^1}\,,
\end{equation}
and construct separate approximations for the $n_l$-independent
coefficients $\Pi^{(2)}_{n_l^0}, \Pi^{(2)}_{n_l^1}$. After discarding
unphysical approximants as described above we obtain 80 approximants for
each $\Pi^{(2)}_{n_l^0}$ and $\Pi^{(2)}_{n_l^1}$. The expressions for
the approximants are quite lengthy and provided as ancillary files. Diagonal
approximants with $n=m$ are generally expected to perform best, so we select the Padé
approximants $[n/m]$ that minimise the distance $|n-m|$ for the following comparison.
This corresponds to $[30/30]$ for $\Pi^{(2)}_{n_l^0}$ and either of $[32/28]$
or $[28/32]$ for $\Pi^{(2)}_{n_l^1}$. Since the two latter approximants are
numerically essentially indistinguishable, we somewhat arbitrarily
select $[32/28]$.

\begin{figure}
  \centering
  \begin{tabular}{cc}
  \includegraphics[width=0.45\linewidth]{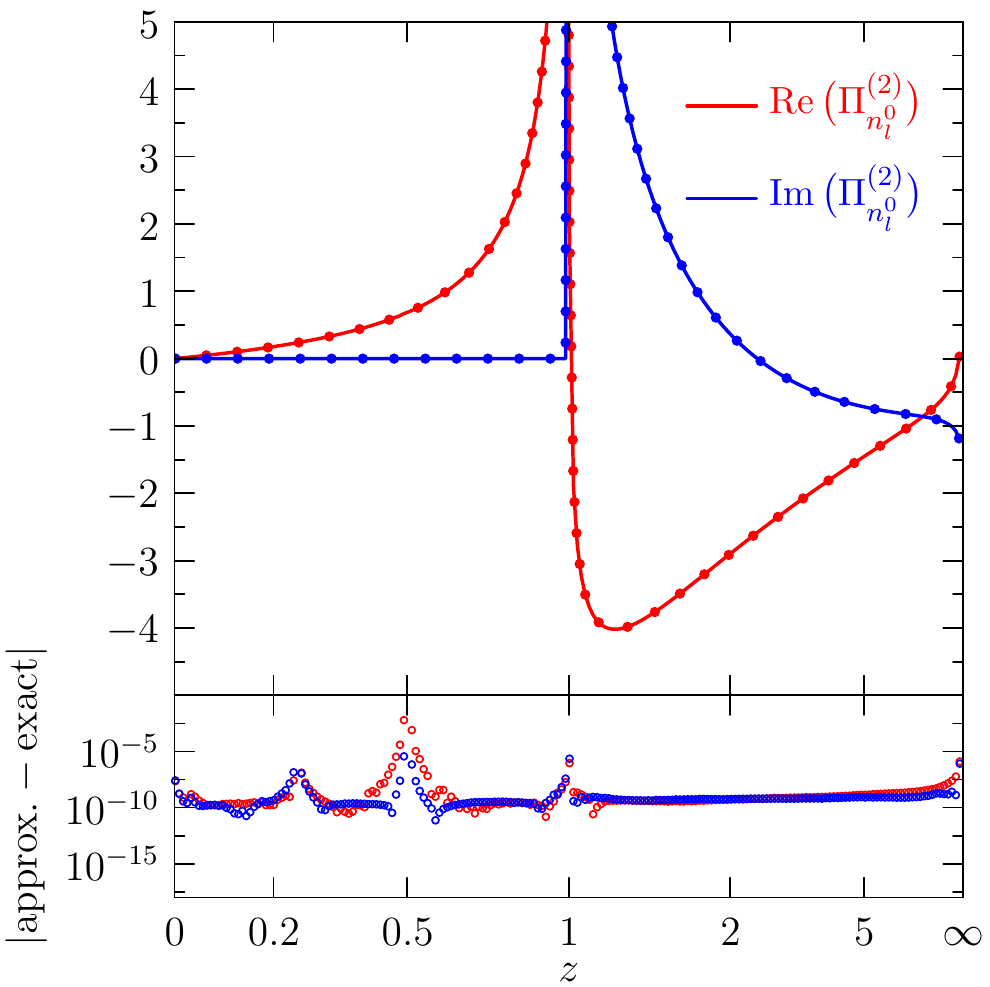}&\includegraphics[width=0.45\linewidth]{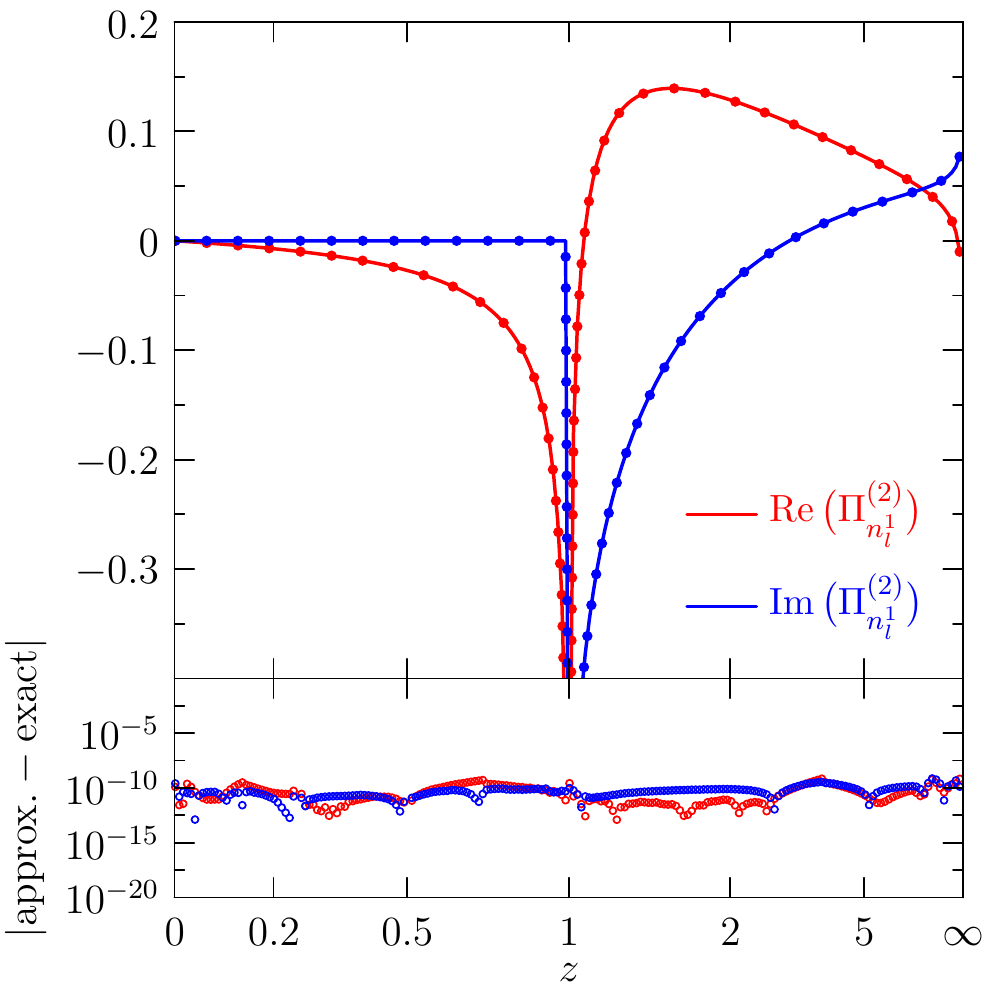}
  \end{tabular}
  \caption{Comparison for $\Pi^{(2)}$ between Padé-based approximation
(dotted) and results obtained by numerically solving differential
equations with a requested absolute error of $10^{-10}$ (solid
lines). The panel on the left shows the corrections without any light
quark flavours, whereas on the right the corrections including a virtual
massless quark loop are considered. Note that the differential equations
contain spurious singularities for $z \in \{0, 0.25, 0.5\}$.}
  \label{fig:pade_cmp}
\end{figure}

In Figure~\ref{fig:pade_cmp} we compare the Padé-based approximants to
the exact result, which we compute as described in
Section~\ref{sec:setup}. For the sake of a clear presentation, we
restrict ourselves to real values of $z$, choosing the physical branch
on the upper complex half-plane for $z>1$. This is implemented in the
numerical evaluation by adding a small imaginary part, i.e. by shifting
the argument $z \to (1 + 10^{-10} i)z$. We solve the differential
equations for 198 values of $z$, which requires about $14$ seconds on a
single core of an Intel Core i5-4200M processor. It should be noted that
the time required for the calculation of a single point increases
greatly in the vicinity of singularities. We find excellent agreement
over the whole kinematic range, including the region around the Coulomb
singularity at $z=1$. In fact, the difference is typically of the order
of the numeric precision requested when solving the differential
equation. We conclude that for all practical purposes the approximation
is indistinguishable from the true result.

With this degree of accuracy, it is also possible to omit a number of
expansion terms in the construction of the approximation while still
retaining agreement with the exact result at the level of
$10^{-10}$. For instance, we find that limiting ourselves to
coefficients $C^{(2)}_n, D^{(2)}_{m,0}$ with $n,m < 22$ does not lead to
a visible increase in the deviation. When omitting further coefficients
the accuracy degrades notably in the region above threshold, e.g. to the
level of $10^{-9}$ for a $[20/20]$ approximant constructed from
coefficients with $n < 20, m < 19$ and $10^{-8}$ for a $[16/15]$
approximant from expansion terms with $n,m < 15$.

\subsection{Comparison to approximation based on Mellin-Barnes transform}
\label{sec:mellin_comp}

\begin{figure}
  \centering
  \includegraphics[width=0.8\linewidth]{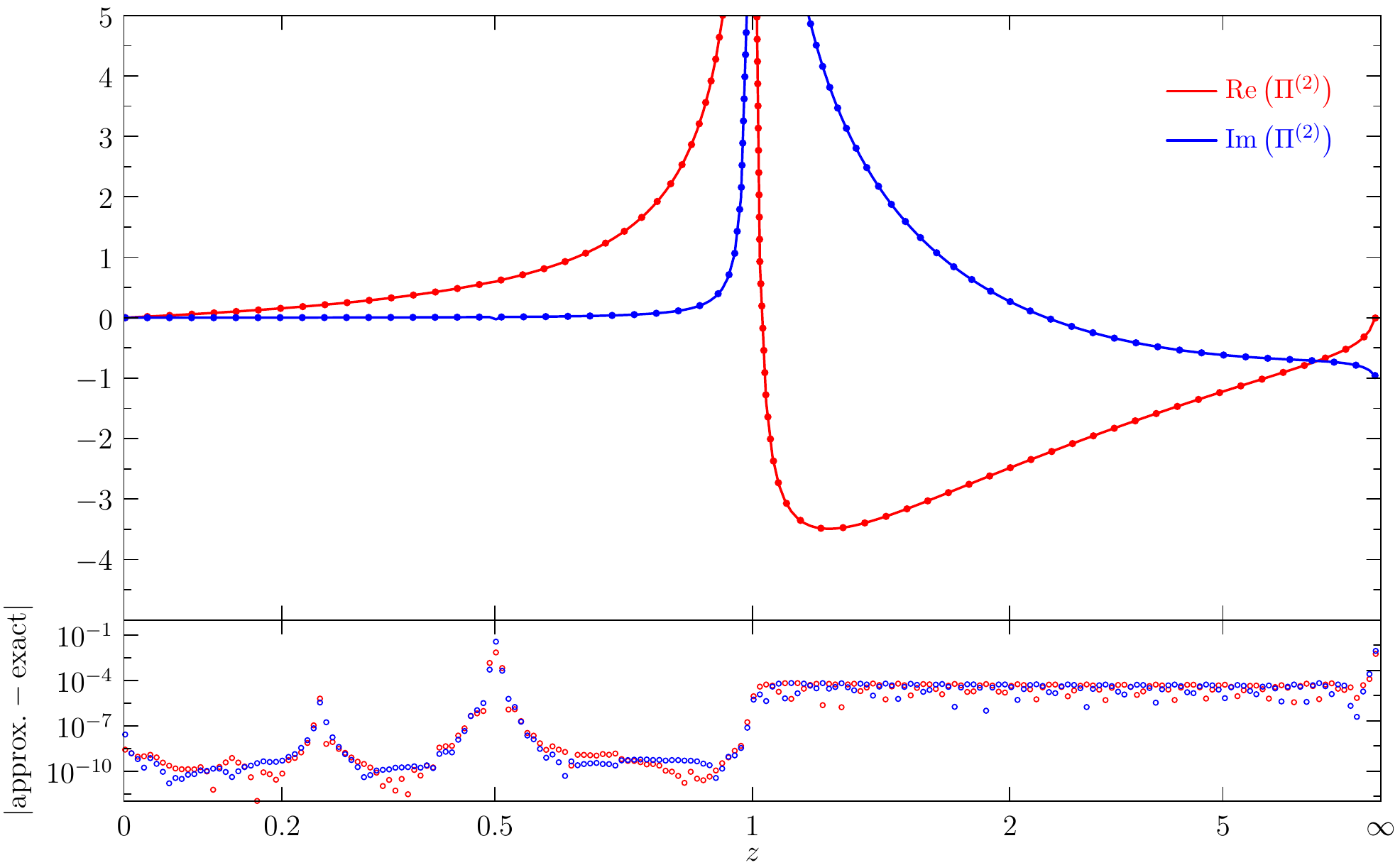}
  \caption{Comparison for $\Pi^{(2)}$ with $n_l=3$ massless quark
    flavours between the approximation of~\cite{Greynat:2010kx}
(dotted) and results obtained by numerically solving differential
equations with a requested absolute error of $10^{-10}$ (solid
lines).}
  \label{fig:mellin_cmp}
\end{figure}

In Figure~\ref{fig:mellin_cmp} we compare the exact result to the
approximation of~\cite{Greynat:2010kx}, which is based on the
Mellin-Barnes transform. In~\cite{Greynat:2010kx}, a flexible number of
$N^*$ coefficients in the low-energy expansion, all known
coefficients in the threshold expansion, and terms up to order $z^{-2}$
in the high-energy expansion are employed in the construction of the
approximation. For the comparison we take into account all $N^* =
30$ low-energy coefficients, but make no attempt at improving the
approximation over what was done in~\cite{Greynat:2010kx}. Similarly to
section~\ref{sec:pade_comp}, we focus on values of $z$ that are close to
the real axis. However, we choose a somewhat larger imaginary part by
shifting $z \to (1 + 0.01 i)z$ in both the approximation and the exact
result. The reason for this is that the expression for the approximant
contains sums of the form $\sum_{n=1}^\infty \tfrac{\log n}{n}(\pm
\omega)^n$, which are difficult to evaluate close to the branch cut
$|\omega|=1$.

As for the Padé-based approximation the agreement in the low-energy
region $z < 1$ is remarkably good. Above the threshold, the difference
is of the order of $10^{-4}$, bigger than for the Padé-based
approximation. We expect that the inclusion of the complete known
high-energy expansion up to $z^{-30}$~\cite{Maier:2011jd} would improve
the precision in this region further.

\section{Low-energy expansion at four loops}

In the following we compare our new analytic result for $C^{(3)}_4$ to
various estimates. A similar comparison at three-loop order using
restricted input in the construction of a Padé-based approximation was
already performed in~\cite{Hoang:2008qy}, where good agreement between
the approximate and exact results for $C^{(2)}_4$ was found.

The low-energy expansion coefficients $C^{(3)}_n$ can be decomposed
according to their colour structure:\footnote{Our notation differs
  slightly from the one employed in e.g.~\cite{Maier:2009fz}, since the latter
  does not generalise well to the purely bosonic corrections.}
\begin{equation}
  \label{eq:C_decomp}
  \begin{split}
    C^{(3)}_n ={}& C_F\*T_F^2\*n_l^2\*C^{(3)}_{ll,n}  + C_F\*T_F^2\*n_l\*n_h\*C^{(3)}_{lh,n} + C_F\*T_F^2\*n_h^2\*C^{(3)}_{hh,n}\\
    &+ C_F\*T_F\*n_l\*\big(C_A\*C^{(3)}_{lA,n} + C_F\*C^{(3)}_{lF,n}\big)
    + C_F\*T_F\*n_h\*\big(C_A\*C^{(3)}_{hA,n} + C_F\*C^{(3)}_{hF,n})\\
    &+ C_F\*\big(C_A^2\*C^{(3)}_{AA,n} + C_A\*C_F\*C^{(3)}_{AF,n} + C_F^2\*C^{(3)}_{FF,n}\big) + \frac{d_{33}^{FF}}{D_F}\*C^{(3)}_{\text{sing},n}\,.
  \end{split}
\end{equation}
As usual, $C_F$ and $C_A$ denote the eigenvalues of the quadratic Casimir
operators in the fundamental and the adjoint representation,
respectively. $T_F$ is the trace normalisation defined by $\tr(T^a T^b)
= T_F \delta^{ab}$, where $T^b, T^b$ are generators of the fundamental
representation. For QCD, the values of these colour factors are $C_F =
4/3, C_A = 3, T_F = 1/2$. The number of quark flavours with mass $m_Q$
is denoted by $n_h$. The remaining factors in Eq.~(\ref{eq:C_decomp})
are the dimension of the fundamental representation $D_F$ and
$d_{33}^{FF} = [\tfrac{1}{2}\tr(T^aT^bT^c + T^a T^c T^b)]^2$. However,
this colour structure only appears in the flavour-singlet
contribution. As already mentioned in Section~\ref{sec:intro}, we will
therefore not consider the coefficient $C^{(3)}_{\text{sing},n}$.

Since the bosonic contribution for $C^{(3)}_2,C^{(3)}_3$ has only been
presented for a SU(3) gauge group in previous
works~\cite{Maier:2008he,Maier:2009fz}, we provide the general colour
decomposition in Appendix~\ref{sec:Cn_res}. Our new result for
$C^{(3)}_4$ reads
\begin{align}
  \label{eq:C_4}
  C^{(3)}_{ll,4} ={}& \frac{111598019584}{113927664375}+\frac{3328}{18711}\*\pi^2\,,\displaybreak[0]\\
  C^{(3)}_{lh,4}
  ={}&\frac{10452332929019}{2149908480000}-\frac{3328}{18711}\*\pi^2-\frac{1868838269}{424673280}\*\zeta_3+\frac{17659747}{637009920}\*\pi^4-\frac{360403}{5308416}\*c_4\,,\displaybreak[0]\\
  C^{(3)}_{hh,4} ={}&\frac{49043275373141}{5764442112000}-\frac{4096}{93555}\*\pi^2-\frac{800398998419}{119558799360}\*\zeta_3\,,\displaybreak[0]\\
  C^{(3)}_{lA,4} ={}&-\frac{1545856136885976983}{597309072998400000}-\frac{1600}{2673}\*\pi^2-\frac{512}{567}\*\pi^2\*\log(2)-\frac{286823413412357}{802632499200}\*\zeta_3\notag\\
&+\frac{1284529483609}{294298583040}\*\pi^4+\frac{512}{2079}\*\pi^2\*\log^2(2)-\frac{3954329}{206438400}\*c_4\,,\displaybreak[0]\\
  C^{(3)}_{lF,4} ={}&\frac{99633942573144089459}{77428953907200000}-\frac{8808368}{4209975}\*\pi^2+\frac{1024}{567}\*\pi^2\*\log(2)-\frac{7582402055503}{990904320}\*\zeta_3\notag\\
&+\frac{4739656702961}{58392576000}\*\pi^4-\frac{1024}{2079}\*\pi^2\*\log^2(2)+\frac{3954329}{103219200}\*c_4\,,\displaybreak[0]\\
  C^{(3)}_{hA,4} ={}&\frac{1101070706845234821395897}{216048997032110161920000}+\frac{28736}{6237}\*\pi^2-\frac{32768}{6237}\*\pi^2\*\log(2)\notag\\
&-\frac{42230989766134848484889}{152415518188437504000}\*\zeta_3+\frac{612247348225991}{143470559232000}\*\pi^4\notag\\
&-\frac{5054901194017}{298896998400}\*c_4-\frac{128}{693}\*\pi^2\*\zeta_3+\frac{2368}{63}\*\zeta_5\,,\displaybreak[0]\\
  C^{(3)}_{hF,4} ={}&-\frac{177940168537927422175447}{31950458005340160000}+\frac{1932448}{601425}\*\pi^2-\frac{8192}{6237}\*\pi^2\*\log(2)\notag\\
&-\frac{2700109017390851879983}{73630685115187200}\*\zeta_3+\frac{284593079466233}{398529331200}\*\pi^4-\frac{16742297834089}{6642155520}\*c_4\,,\displaybreak[0]\\
  C^{(3)}_{AA,4} ={}&\frac{1909337920002502630087183687}{1213920785999816294400000}+\frac{15640}{18711}\*\pi^2+\frac{14720}{6237}\*\pi^2\*\log(2)\notag\\
&-\frac{14370134990138593178785411}{1998223516049080320000}\*\zeta_3+\frac{1402361362646965369001}{7785286426165248000}\*\pi^4\notag\\
&-\frac{128}{189}\*\pi^2\*\log^2(2)-\frac{68904556891714010393}{64877386884710400}\*c_4-\frac{14659863890116}{29462808375}\*\pi^4\*\log(2)\notag\\
&-\frac{272}{231}\*\pi^2\*\zeta_3+\frac{39155511739826}{654729075}\*\zeta_5+\frac{4512791972672}{29462808375}\*c_5\,,\displaybreak[0]\\
  C^{(3)}_{AF,4} ={}&-\frac{10166890739766552788195414539}{1416240916999785676800000}-\frac{1401824}{601425}\*\pi^2+\frac{473488}{127575}\*\pi^2\*\log(2)\notag\\
&+\frac{1816432535076972153588341}{25618250205757440000}\*\zeta_3-\frac{22065121460722643973581}{19463216065413120000}\*\pi^4\notag\\
&+\frac{256}{189}\*\pi^2\*\log^2(2)+\frac{150856650327341723953}{32438693442355200}\*c_4+\frac{20846843102608}{9820936125}\*\pi^4\*\log(2)\notag\\
&+\frac{1216}{693}\*\pi^2\*\zeta_3-\frac{15372485043544}{59520825}\*\zeta_5-\frac{5927657182976}{9820936125}\*c_5\,,\displaybreak[0]\\
  C^{(3)}_{FF,4} ={}&-\frac{91078971803776091210773}{74382401102929920000}+\frac{218452}{25515}\*\pi^2-\frac{23664736}{1403325}\*\pi^2\*\log(2)\notag\\
&-\frac{764847650806674196397063}{10407414146088960000}\*\zeta_3+\frac{100112202889759409717}{76028187755520000}\*\pi^4\notag\\
&-\frac{2048802335890692839}{405483668029440}\*c_4-\frac{9491657579312}{4208972625}\*\pi^4\*\log(2)\notag\\
&+\frac{64}{693}\*\pi^2\*\zeta_3+\frac{171651515018024}{654729075}\*\zeta_5+\frac{17514775207168}{29462808375}\*c_5\,,
\end{align}
where $\zeta_n = \sum_{k=1}^\infty \tfrac{1}{k^n}$ denote values of the Riemann $\zeta$
function, the auxiliary constants $c_4, c_5$ are given by
\begin{align}
  \label{eq:c_45}
c_4 ={}& 24\*\Li_4\big(\tfrac{1}{2}\big) + \log^4(2) - \pi^2
         \log^2(2)\,,\\
c_5 ={}& -360\*\Li_5\big(\tfrac{1}{2}\big) + 3\*\log^5(2) - 5\*\pi^2 \log^3(2)\,,
\end{align}
and $\Li_n(\tfrac{1}{2}) = \sum_{k=1}^\infty \tfrac{1}{2^kk^n}$ are
values of polylogarithm functions. The corresponding results for the
coefficients in the \MS{} scheme are given in
Appendix~\ref{sec:Cn_MS}. The expressions in both schemes are also
available in computer-readable form as ancillary files.

In Table~\ref{tab:C4_comp} we compare the numerical values for QCD with
$n_l = 3,4,5$ to the estimates obtained
in~\cite{Hoang:2008qy,Kiyo:2009gb,Greynat:2011zp}. We find excellent
agreement, especially for the predictions from~\cite{Kiyo:2009gb}. In
fact, the true approximation error of~\cite{Kiyo:2009gb} seems to be
almost an order of magnitude less than estimated.

\begin{table}
  \centering
  \begin{tabular}{c@{\qquad}c}
  \begin{tabular}{lr@{.}lr@{.}lr@{.}l}
    \toprule
\multirow{2}{*}{$n_l$} & \multicolumn{6}{c}{$C^{(3)}_4$} \\
    \cmidrule(lr){2-7}
                       &\multicolumn{2}{c}{Ref.~\cite{Greynat:2011zp}}&\multicolumn{2}{c}{Ref.~\cite{Kiyo:2009gb}}&\multicolumn{2}{c}{exact}\\
    \midrule
    3 & 382&7(5)                & 383&073(11) & 383&075\\
    4 & 339&7(5)                & 339&913(10) & 339&913\\
    5 & \multicolumn{2}{c}{---} & 298&576(9)  & 298&575\\
    \bottomrule
  \end{tabular}
&
  \begin{tabular}{lr@{.}lr@{.}lr@{.}l}
    \toprule
\multirow{2}{*}{$n_l$} & \multicolumn{6}{c}{$\bar{C}^{(3)}_4$} \\
    \cmidrule(lr){2-7}
                       &\multicolumn{2}{c}{Ref.~\cite{Hoang:2008qy}} &\multicolumn{2}{c}{Ref.~\cite{Kiyo:2009gb}}&\multicolumn{2}{c}{exact}\\
    \midrule
    3 & -4&238(1171)             & -3&349(11) &-3&348\\
    4 & -1&935(1201)             & -1&386(10) &-1&386\\
    5 &  \multicolumn{2}{c}{---} &  0&471(9)  &0&471\\
    \bottomrule
  \end{tabular}
  \end{tabular}
  \caption{Comparison for $C^{(3)}_4$ between predictions
    from~\cite{Hoang:2008qy,Kiyo:2009gb,Greynat:2011zp} and the exact analytic result for different
    numbers of light quark flavours. $\bar{C}^{(3)}_4$ is the coefficient in
    the \MS{} scheme. We have refrained from converting the results
    from~\cite{Hoang:2008qy} and ~\cite{Greynat:2011zp} to different schemes.}
  \label{tab:C4_comp}
\end{table}

\section{Conclusion}
\label{sec:conclusion}

We have tested the quality of three- and four-loop approximations for
the quark contribution to the vacuum polarisation. To this end, we have
calculated the three-loop contribution numerically, finding almost perfect
agreement with a newly constructed Padé-based approximation and
very good agreement with an approximation from~\cite{Greynat:2010kx}. At
four loops, we have computed analytically the fourth term in the
low-energy expansion, which is also relevant for relativistic sum-rule
determinations of the charm- and bottom-quark masses. We find
excellent agreement with the Padé-based prediction~\cite{Kiyo:2009gb}, well
within the error estimate. Within their errors, the less precise estimates
from~\cite{Hoang:2008qy} and~\cite{Greynat:2011zp} also agree well with
the exact result.

\section*{Acknowledgements}

We thank T.~Rauh for providing the approximant
of~\cite{Greynat:2010kx} in electronic form and M.~Steinhauser for
proofreading and suggesting a number of improvements. A.~M. was
supported by a European Union COFUND/Durham Junior Research Fellowship
under EU grant agreement number 267209. P.~M. was supported in part by
the EU Network HIGGSTOOLS PITN-GA-2012-316704.

\appendix

\section{Results for the low-energy expansion at four loops}
\label{sec:Cn_res}

\subsection{Coefficients in the on-shell scheme}
\label{sec:Cn_OS}

In the following, we show the four-loop coefficients $C_n^{(3)}$ with
$n\in \{1,2,3\}$ in a general gauge group, using the decomposition
Eq.~(\ref{eq:C_decomp}).
\begin{align}
  C^{(3)}_{ll,1} ={}&\frac{12670}{6561}+\frac{104}{405}\*\pi^2
\,,\displaybreak[0]\\
  C^{(3)}_{lh,1} ={}&\frac{222985}{52488}-\frac{104}{405}\*\pi^2-\frac{9625}{3888}\*\zeta_3+\frac{1421}{58320}\*\pi^4-\frac{29}{486}\*c_4
\,,\displaybreak[0]\\
  C^{(3)}_{hh,1} ={}&\frac{83971}{19683}-\frac{128}{2025}\*\pi^2-\frac{1291}{486}\*\zeta_3
\,,\displaybreak[0]\\
  C^{(3)}_{lA,1} ={}&-\frac{10827311}{839808}-\frac{70}{81}\*\pi^2-\frac{63301}{6912}\*\zeta_3-\frac{176}{135}\*\pi^2\*\log(2)\notag\\
&+\frac{92437}{933120}\*\pi^4+\frac{16}{45}\*\pi^2\*\log^2(2)-\frac{37}{7776}\*c_4
\,,\displaybreak[0]\\
  C^{(3)}_{lF,1} ={}&-\frac{96275}{15552}-\frac{2584}{1215}\*\pi^2+\frac{352}{135}\*\pi^2\*\log(2)-\frac{812179}{31104}\*\zeta_3\notag\\
&+\frac{32815}{93312}\*\pi^4-\frac{32}{45}\*\pi^2\*\log^2(2)+\frac{37}{3888}\*c_4
\,,\displaybreak[0]\\
  C^{(3)}_{hA,1} ={}&-\frac{65230603633}{2829103200}+\frac{898}{135}\*\pi^2-\frac{1024}{135}\*\pi^2\*\log(2)-\frac{3628884481}{209563200}\*\zeta_3\notag\\
&+\frac{189821}{510300}\*\pi^4-\frac{33511}{34020}\*c_4-\frac{4}{15}\*\pi^2\*\zeta_3+\frac{100}{27}\*\zeta_5
\,,\displaybreak[0]\\
  C^{(3)}_{hF,1} ={}&-\frac{457939}{14700}+\frac{3464}{1215}\*\pi^2-\frac{256}{135}\*\pi^2\*\log(2)-\frac{77255063}{1190700}\*\zeta_3+\frac{473237}{340200}\*\pi^4-\frac{30853}{5670}\*c_4
\,,\displaybreak[0]\\
  C^{(3)}_{AA,1} ={}&\frac{1600078157}{83980800}+\frac{391}{324}\*\pi^2+\frac{92}{27}\*\pi^2\*\log(2)-\frac{317785087}{2073600}\*\zeta_3+\frac{366629309}{130636800}\*\pi^4\notag\\
&-\frac{44}{45}\*\pi^2\*\log^2(2)-\frac{13924493}{1088640}\*c_4-\frac{83311}{34020}\*\pi^4\*\log(2)-\frac{17}{10}\*\pi^2\*\zeta_3+\frac{18091}{56}\*\zeta_5+\frac{6542}{8505}\*c_5
\,,\displaybreak[0]\\
  C^{(3)}_{AF,1} ={}&-\frac{8573321}{1209600}-\frac{1201}{486}\*\pi^2+\frac{1084}{405}\*\pi^2\*\log(2)+\frac{12340560719}{21772800}\*\zeta_3-\frac{21178249}{2419200}\*\pi^4\notag\\
&+\frac{88}{45}\*\pi^2\*\log^2(2)+\frac{6093697}{181440}\*c_4+\frac{2935}{378}\*\pi^4\*\log(2)+\frac{38}{15}\*\pi^2\*\zeta_3-\frac{788465}{756}\*\zeta_5-\frac{2204}{945}\*c_5
\,,\displaybreak[0]\\
  C^{(3)}_{FF,1} ={}&\frac{3562169}{680400}+\frac{7297}{810}\*\pi^2-\frac{7688}{405}\*\pi^2\*\log(2)-\frac{69694097}{302400}\*\zeta_3+\frac{66975707}{16329600}\*\pi^4\notag\\
&-\frac{2178299}{136080}\*c_4-\frac{48764}{8505}\*\pi^4\*\log(2)+\frac{2}{15}\*\pi^2\*\zeta_3+\frac{135724}{189}\*\zeta_5+\frac{13504}{8505}\*c_5\,,
\end{align}
\begin{align}
  C^{(3)}_{ll,2} ={}&\frac{3718639}{2733750}+\frac{208}{945}\*\pi^2
\,,\displaybreak[0]\\
  C^{(3)}_{lh,2} ={}&\frac{192528671}{44789760}-\frac{208}{945}\*\pi^2-\frac{10033247}{3317760}\*\zeta_3+\frac{99421}{4976640}\*\pi^4-\frac{2029}{41472}\*c_4
\,,\displaybreak[0]\\
  C^{(3)}_{hh,2} ={}&\frac{3284183491}{646652160}-\frac{256}{4725}\*\pi^2-\frac{19669747}{5322240}\*\zeta_3
\,,\displaybreak[0]\\
  C^{(3)}_{lA,2} ={}&-\frac{162305982733}{16796160000}-\frac{20}{27}\*\pi^2-\frac{352}{315}\*\pi^2\*\log(2)-\frac{15414971}{614400}\*\zeta_3\notag\\
&+\frac{159124087}{522547200}\*\pi^4+\frac{32}{105}\*\pi^2\*\log^2(2)-\frac{11233}{622080}\*c_4
\,,\displaybreak[0]\\
  C^{(3)}_{lF,2} ={}&\frac{37083738217}{1679616000}-\frac{4316}{2025}\*\pi^2+\frac{704}{315}\*\pi^2\*\log(2)-\frac{5375180501}{24883200}\*\zeta_3\notag\\
&+\frac{124955317}{52254720}\*\pi^4-\frac{64}{105}\*\pi^2\*\log^2(2)+\frac{11233}{311040}\*c_4
\,,\displaybreak[0]\\
  C^{(3)}_{hA,2} ={}&-\frac{116711876411987}{5649153269760}+\frac{1796}{315}\*\pi^2-\frac{28672703485597}{697426329600}\*\zeta_3-\frac{2048}{315}\*\pi^2\*\log(2)\notag\\
&+\frac{214047541}{278691840}\*\pi^4-\frac{30416201}{11612160}\*c_4-\frac{8}{35}\*\pi^2\*\zeta_3+\frac{62}{9}\*\zeta_5
\,,\displaybreak[0]\\
  C^{(3)}_{hF,2} ={}&-\frac{43144263820961}{271593907200}+\frac{43384}{14175}\*\pi^2-\frac{5763324918049}{6706022400}\*\zeta_3-\frac{512}{315}\*\pi^2\*\log(2)\notag\\
&+\frac{2208846791}{130636800}\*\pi^4-\frac{130829911}{2177280}\*c_4
\,,\displaybreak[0]\\
  C^{(3)}_{AA,2} ={}&\frac{466388623105831}{36212520960000}+\frac{391}{378}\*\pi^2+\frac{184}{63}\*\pi^2\*\log(2)-\frac{140086837646759}{211341312000}\*\zeta_3\notag\\
&+\frac{44385041477}{3335904000}\*\pi^4-\frac{88}{105}\*\pi^2\*\log^2(2)-\frac{12059516453}{194594400}\*c_4\notag\\
&-\frac{117177241}{8108100}\*\pi^4\*\log(2)-\frac{51}{35}\*\pi^2\*\zeta_3+\frac{650792267}{360360}\*\zeta_5+\frac{9162968}{2027025}\*c_5
\,,\displaybreak[0]\\
  C^{(3)}_{AF,2} ={}&-\frac{1533334898954081}{5884534656000}-\frac{34397}{14175}\*\pi^2+\frac{15212}{4725}\*\pi^2\*\log(2)+\frac{148076196562157}{67060224000}\*\zeta_3\notag\\
&-\frac{47361359145059}{1494484992000}\*\pi^4+\frac{176}{105}\*\pi^2\*\log^2(2)+\frac{1604148237623}{12454041600}\*c_4\notag\\
&+\frac{641695511}{12162150}\*\pi^4\*\log(2)+\frac{76}{35}\*\pi^2\*\zeta_3-\frac{1210376569}{180180}\*\zeta_5-\frac{94165256}{6081075}\*c_5
\,,\displaybreak[0]\\
  C^{(3)}_{FF,2} ={}&\frac{94330906317547}{871782912000}+\frac{1678}{189}\*\pi^2-\frac{12232}{675}\*\pi^2\*\log(2)-\frac{112244692092317}{1743565824000}\*\zeta_3\notag\\
&-\frac{427149037853}{747242496000}\*\pi^4-\frac{60530131639}{6227020800}\*c_4\notag\\
&-\frac{290163788}{6081075}\*\pi^4\*\log(2)+\frac{4}{35}\*\pi^2\*\zeta_3+\frac{28631674}{5005}\*\zeta_5+\frac{78374896}{6081075}\*c_5
\end{align}
\begin{align}
  C^{(3)}_{ll,3} ={}&\frac{55769012272}{49228003125}+\frac{1664}{8505}\*\pi^2
\,,\displaybreak[0]\\
  C^{(3)}_{lh,3} ={}&\frac{133069561477}{29393280000}-\frac{1664}{8505}\*\pi^2-\frac{9137231}{2488320}\*\zeta_3+\frac{17297}{746496}\*\pi^4-\frac{1765}{31104}\*c_4
\,,\displaybreak[0]\\
  C^{(3)}_{hh,3} ={}&\frac{81866930683}{12609717120}-\frac{2048}{42525}\*\pi^2-\frac{7731286469}{1556755200}\*\zeta_3
\,,\displaybreak[0]\\
  C^{(3)}_{lA,3} ={}&-\frac{49195270842508337}{6452412825600000}-\frac{160}{243}\*\pi^2-\frac{2816}{2835}\*\pi^2\*\log(2)-\frac{1702438681003}{18579456000}\*\zeta_3\notag\\
&+\frac{31331237083}{27869184000}\*\pi^4+\frac{256}{945}\*\pi^2\*\log^2(2)-\frac{4617931}{232243200}\*c_4
\,,\displaybreak[0]\\
  C^{(3)}_{lF,3} ={}&\frac{1330122477499829}{6145155072000}-\frac{349544}{165375}\*\pi^2+\frac{5632}{2835}\*\pi^2\*\log(2)-\frac{9692093720699}{7225344000}\*\zeta_3\notag\\
&+\frac{198998967077}{13934592000}\*\pi^4-\frac{512}{945}\*\pi^2\*\log^2(2)+\frac{4617931}{116121600}\*c_4
\,,\displaybreak[0]\\
  C^{(3)}_{hA,3} ={}&-\frac{89146729206385547629}{5854170457175040000}+\frac{14368}{2835}\*\pi^2-\frac{16384}{2835}\*\pi^2\*\log(2)\notag\\
&-\frac{43342214888270310611}{433642256087040000}\*\zeta_3+\frac{123542892287}{74511360000}\*\pi^4\notag\\
&-\frac{778807933}{124185600}\*c_4-\frac{64}{315}\*\pi^2\*\zeta_3+\frac{4768}{315}\*\zeta_5
\,,\displaybreak[0]\\
  C^{(3)}_{hF,3} ={}&-\frac{2392972916257093}{2489610816000}+\frac{4701392}{1488375}\*\pi^2-\frac{4096}{2835}\*\pi^2\*\log(2)-\frac{377287031234107}{61471872000}\*\zeta_3\notag\\
&+\frac{95566793477}{798336000}\*\pi^4-\frac{1875259367}{4435200}\*c_4
\,,\displaybreak[0]\\
  C^{(3)}_{AA,3} ={}&\frac{911077165600708705799}{11072340408729600000}+\frac{1564}{1701}\*\pi^2+\frac{1472}{567}\*\pi^2\*\log(2)-\frac{6083812406329392817}{2107777351680000}\*\zeta_3\notag\\
&+\frac{83515526788350341}{1354999726080000}\*\pi^4-\frac{704}{945}\*\pi^2\*\log^2(2)-\frac{671492238494897}{2258332876800}\*c_4\notag\\
&-\frac{44413927637}{516891375}\*\pi^4\*\log(2)-\frac{136}{105}\*\pi^2\*\zeta_3+\frac{240485544239}{22972950}\*\zeta_5+\frac{13771367704}{516891375}\*c_5
\,,\displaybreak[0]\\
  C^{(3)}_{AF,3} ={}&-\frac{654109650631266719803}{384142422343680000}-\frac{392944}{165375}\*\pi^2+\frac{1745896}{496125}\*\pi^2\*\log(2)\notag\\
&+\frac{103053230050290459193}{9484998082560000}\*\zeta_3-\frac{322882823584776463}{2032499589120000}\*\pi^4\notag\\
&+\frac{1408}{945}\*\pi^2\*\log^2(2)+\frac{2266030736034323}{3387499315200}\*c_4+\frac{180265104542}{516891375}\*\pi^4\*\log(2)\notag\\
&+\frac{608}{315}\*\pi^2\*\zeta_3-\frac{164899159109}{3828825}\*\zeta_5-\frac{7402036192}{73841625}\*c_5
\,,\displaybreak[0]\\
  C^{(3)}_{FF,3} ={}&\frac{300002162759308069}{375139084320000}+\frac{10674}{1225}\*\pi^2-\frac{2881264}{165375}\*\pi^2\*\log(2)-\frac{1531450738927589}{842063040000}\*\zeta_3\notag\\
&+\frac{176076905389817}{7939451520000}\*\pi^4-\frac{245658223193}{1654052400}\*c_4-\frac{182874498536}{516891375}\*\pi^4\*\log(2)\notag\\
&+\frac{32}{315}\*\pi^2\*\zeta_3+\frac{476561040896}{11486475}\*\zeta_5+\frac{48543035872}{516891375}\*c_5
\end{align}

\subsection{Coefficients in the \MS{} scheme}
\label{sec:Cn_MS}

Renormalising the heavy-quark mass in the \MS{} instead of the on-shell
scheme we obtain the low-energy expansion
\begin{equation}
  \label{eq:P3_le} \Pi^{(3)}(q^2) = \sum_{n=1}^\infty \bar{C}_n^{(3)} \bigg(\frac{q^2}{4\bar{m}^2_Q}\bigg)^n\,,
\end{equation}
where $\bar{m}_Q$ now denotes the \MS{} mass~\cite{Tarasov:1982gk,Chetyrkin:1997dh,Vermaseren:1997fq,Baikov:2014qja,Luthe:2016xec,Baikov:2017ujl} at the scale $\mu =
\bar{m}_Q$. The analytic results are
\begin{equation}
  \begin{split}
    \bar{C}^{(3)}_n ={}& C_F\*T_F^2\*n_l^2\*\bar{C}^{(3)}_{ll,n}  + C_F\*T_F^2\*n_l\*n_h\*\bar{C}^{(3)}_{lh,n} + C_F\*T_F^2\*n_h^2\*\bar{C}^{(3)}_{hh,n}\\
    &+ C_F\*T_F\*n_l\*\big(C_A\*\bar{C}^{(3)}_{lA,n} +
    C_F\*\bar{C}^{(3)}_{lF,n}\big)
    + C_F\*T_F\*n_h\*\big(C_A\*\bar{C}^{(3)}_{hA,n} + C_F\*\bar{C}^{(3)}_{hF,n})\\
    &+ C_F\*\big(C_A^2\*\bar{C}^{(3)}_{AA,n} +
    C_A\*C_F\*\bar{C}^{(3)}_{AF,n} + C_F^2\*\bar{C}^{(3)}_{FF,n}\big) +
    \frac{d_{33}^{FF}}{D_F}\*\bar{C}^{(3)}_{\text{sing},n}\,,
  \end{split}
\end{equation}
\begin{align}
\bar{C}^{(3)}_{ll,1} ={}&\frac{42173}{32805}-\frac{112}{135}\*\zeta_3
\,,\displaybreak[0]\\
  \bar{C}^{(3)}_{lh,1}
  ={}&\frac{262877}{262440}-\frac{38909}{19440}\*\zeta_3+\frac{1421}{58320}\*\pi^4-\frac{29}{486}\*c_4
\,,\displaybreak[0]\\
  \bar{C}^{(3)}_{hh,1} ={}&\frac{163868}{98415}-\frac{3287}{2430}\*\zeta_3
\,,\displaybreak[0]\\
  \bar{C}^{(3)}_{lA,1} ={}&-\frac{13377067}{4199040}-\frac{270937}{34560}\*\zeta_3+\frac{549737}{4665600}\*\pi^4-\frac{4793}{38880}\*c_4
\,,\displaybreak[0]\\
  \bar{C}^{(3)}_{lF,1} ={}&\frac{168257}{77760}-\frac{660115}{31104}\*\zeta_3+\frac{546199}{2332800}\*\pi^4+\frac{4793}{19440}\*c_4
\,,\displaybreak[0]\\
  \bar{C}^{(3)}_{hA,1} ={}&-\frac{8974914913}{2829103200}-\frac{3967290241}{209563200}\*\zeta_3+\frac{146477}{510300}\*\pi^4-\frac{37543}{34020}\*c_4+\frac{64}{27}\*\zeta_5
\,,\displaybreak[0]\\
  \bar{C}^{(3)}_{hF,1} ={}&-\frac{53113237}{3572100}-\frac{42551329}{595350}\*\zeta_3+\frac{503813}{340200}\*\pi^4-\frac{29509}{5670}\*c_4
\,,\displaybreak[0]\\
  \bar{C}^{(3)}_{AA,1} ={}&-\frac{304386643}{83980800}-\frac{297156607}{2073600}\*\zeta_3+\frac{352194749}{130636800}\*\pi^4-\frac{13569677}{1088640}\*c_4\notag\\
&-\frac{83311}{34020}\*\pi^4\*\log(2)+\frac{53545}{168}\*\zeta_5+\frac{6542}{8505}\*c_5
\,,\displaybreak[0]\\
  \bar{C}^{(3)}_{AF,1} ={}&-\frac{103500329}{10886400}+\frac{11824588199}{21772800}\*\zeta_3-\frac{61205147}{7257600}\*\pi^4+\frac{6072193}{181440}\*c_4\notag\\
&+\frac{2935}{378}\*\pi^4\*\log(2)-\frac{783929}{756}\*\zeta_5-\frac{2204}{945}\*c_5
\,,\displaybreak[0]\\
  \bar{C}^{(3)}_{FF,1} ={}&\frac{4124201}{170100}-\frac{81560567}{302400}\*\zeta_3+\frac{66249947}{16329600}\*\pi^4-\frac{2323451}{136080}\*c_4\notag\\
&-\frac{48764}{8505}\*\pi^4\*\log(2)+\frac{135976}{189}\*\zeta_5+\frac{13504}{8505}\*c_5
\,,
\end{align}
\begin{align}
  \bar{C}^{(3)}_{ll,2} ={}&\frac{15441973}{19136250}-\frac{32}{45}\*\zeta_3
\,,\displaybreak[0]\\
  \bar{C}^{(3)}_{lh,2}
  ={}&\frac{95040709}{62705664}-\frac{12159109}{4644864}\*\zeta_3+\frac{99421}{4976640}\*\pi^4-\frac{2029}{41472}\*c_4
\,,\displaybreak[0]\\
  \bar{C}^{(3)}_{hh,2} ={}&\frac{1842464707}{646652160}-\frac{2744471}{1064448}\*\zeta_3
\,,\displaybreak[0]\\
  \bar{C}^{(3)}_{lA,2} ={}&-\frac{22559166733}{16796160000}-\frac{309132631}{12902400}\*\zeta_3+\frac{167529079}{522547200}\*\pi^4-\frac{520999}{4354560}\*c_4
\,,\displaybreak[0]\\
  \bar{C}^{(3)}_{lF,2} ={}&\frac{357543003871}{11757312000}-\frac{36896356307}{174182400}\*\zeta_3+\frac{598455689}{261273600}\*\pi^4+\frac{520999}{2177280}\*c_4
\,,\displaybreak[0]\\
  \bar{C}^{(3)}_{hA,2} ={}&-\frac{20427854209619}{5649153269760}-\frac{29638030087837}{697426329600}\*\zeta_3\notag\\
&+\frac{968787977}{1393459200}\*\pi^4-\frac{31595849}{11612160}\*c_4+\frac{362}{63}\*\zeta_5
\,,\displaybreak[0]\\
  \bar{C}^{(3)}_{hF,2} ={}&-\frac{37320009196157}{271593907200}-\frac{5811074101069}{6706022400}\*\zeta_3+\frac{2218910663}{130636800}\*\pi^4-\frac{130387543}{2177280}\*c_4
\,,\displaybreak[0]\\
  \bar{C}^{(3)}_{AA,2} ={}&-\frac{237501566974169}{36212520960000}-\frac{138284733633959}{211341312000}\*\zeta_3+\frac{308483709539}{23351328000}\*\pi^4-\frac{1715021939}{27799200}\*c_4\notag\\
&-\frac{117177241}{8108100}\*\pi^4\*\log(2)+\frac{649453787}{360360}\*\zeta_5+\frac{9162968}{2027025}\*c_5
\,,\displaybreak[0]\\
  \bar{C}^{(3)}_{AF,2} ={}&-\frac{254200422088057}{1176906931200}+\frac{143639759446277}{67060224000}\*\zeta_3-\frac{6707168298437}{213497856000}\*\pi^4\notag\\
&+\frac{1602883065143}{12454041600}\*c_4+\frac{641695511}{12162150}\*\pi^4\*\log(2)-\frac{1209449929}{180180}\*\zeta_5-\frac{94165256}{6081075}\*c_5
\,,\displaybreak[0]\\
  \bar{C}^{(3)}_{FF,2} ={}&\frac{1560916977924001}{2615348736000}-\frac{866743989484157}{1743565824000}\*\zeta_3-\frac{455615418653}{747242496000}\*\pi^4-\frac{66223407799}{6227020800}\*c_4\notag\\
&-\frac{290163788}{6081075}\*\pi^4\*\log(2)+\frac{28637394}{5005}\*\zeta_5+\frac{78374896}{6081075}\*c_5
\,,
\end{align}
\begin{align}
  \bar{C}^{(3)}_{ll,3} ={}&\frac{31556642272}{49228003125}-\frac{256}{405}\*\zeta_3
\,,\displaybreak[0]\\
  \bar{C}^{(3)}_{lh,3} ={}&\frac{60361465477}{29393280000}-\frac{57669161}{17418240}\*\zeta_3+\frac{17297}{746496}\*\pi^4-\frac{1765}{31104}\*c_4
\,,\displaybreak[0]\\
  \bar{C}^{(3)}_{hh,3} ={}&\frac{56877138427}{12609717120}-\frac{6184964549}{1556755200}\*\zeta_3
\,,\displaybreak[0]\\
  \bar{C}^{(3)}_{lA,3} ={}&-\frac{1475149211788337}{6452412825600000}-\frac{561258009401}{6193152000}\*\zeta_3+\frac{1510937903}{1327104000}\*\pi^4-\frac{8529817}{77414400}\*c_4
\,,\displaybreak[0]\\
  \bar{C}^{(3)}_{lF,3} ={}&\frac{983812946922223}{4389396480000}-\frac{28995540810097}{21676032000}\*\zeta_3+\frac{21972351293}{1548288000}\*\pi^4+\frac{8529817}{38707200}\*c_4
\,,\displaybreak[0]\\
  \bar{C}^{(3)}_{hA,3} ={}&-\frac{454880458419083629}{5854170457175040000}-\frac{43875740175477222611}{433642256087040000}\*\zeta_3\notag\\
&+\frac{1068488091383}{670602240000}\*\pi^4-\frac{7110196837}{1117670400}\*c_4+\frac{4448}{315}\*\zeta_5
\,,\displaybreak[0]\\
  \bar{C}^{(3)}_{hF,3} ={}&-\frac{2327115263308753}{2489610816000}-\frac{377837317054807}{61471872000}\*\zeta_3\notag\\
&+\frac{286864384271}{2395008000}\*\pi^4-\frac{16870125343}{39916800}\*c_4
\,,\displaybreak[0]\\
  \bar{C}^{(3)}_{AA,3} ={}&\frac{719769197139499105799}{11072340408729600000}-\frac{18203509261598866451}{6323332055040000}\*\zeta_3\notag\\
&+\frac{3971497861375921}{64523796480000}\*\pi^4-\frac{670931439078577}{2258332876800}\*c_4\notag\\
&-\frac{44413927637}{516891375}\*\pi^4\*\log(2)+\frac{240409697039}{22972950}\*\zeta_5+\frac{13771367704}{516891375}\*c_5
\,,\displaybreak[0]\\
  \bar{C}^{(3)}_{AF,3} ={}&-\frac{732946204901779682921}{537799391281152000}+\frac{100087308092213632873}{9484998082560000}\*\zeta_3\notag\\
&-\frac{46055107339265209}{290357084160000}\*\pi^4+\frac{2265724845443603}{3387499315200}\*c_4\notag\\
&+\frac{180265104542}{516891375}\*\pi^4\*\log(2)-\frac{164881655909}{3828825}\*\zeta_5-\frac{7402036192}{73841625}\*c_5
\,,\displaybreak[0]\\
  \bar{C}^{(3)}_{FF,3} ={}&\frac{64791939072296064833}{12004450698240000}-\frac{152847202936444153}{26946017280000}\*\zeta_3\notag\\
&+\frac{175808056237817}{7939451520000}\*\pi^4-\frac{247002468953}{1654052400}\*c_4\notag\\
&-\frac{182874498536}{516891375}\*\pi^4\*\log(2)+\frac{476572709696}{11486475}\*\zeta_5+\frac{48543035872}{516891375}\*c_5
\,,
\end{align}
\begin{align}
  \bar{C}^{(3)}_{ll,4} ={}&\frac{667234795424}{1253204308125}-\frac{512}{891}\*\zeta_3
  \,,\displaybreak[0]\\
  \bar{C}^{(3)}_{lh,4}
  ={}&\frac{432564184014463}{165542952960000}-\frac{44387709491}{10899947520}\*\zeta_3+\frac{17659747}{637009920}\*\pi^4-\frac{360403}{5308416}\*c_4\,,\displaybreak[0]\\
  \bar{C}^{(3)}_{hh,4} ={}&\frac{270605350139987}{40351094784000}-\frac{692437613459}{119558799360}\*\zeta_3
\,,\displaybreak[0]\\
  \bar{C}^{(3)}_{lA,4} ={}&\frac{2470070982166823017}{597309072998400000}-\frac{3146994417526327}{8828957491200}\*\zeta_3\notag\\
&+\frac{1288354688857}{294298583040}\*\pi^4-\frac{2069200171}{20437401600}\*c_4\,,\displaybreak[0]\\
  \bar{C}^{(3)}_{lF,4} ={}&\frac{367248441428202521083}{283906164326400000}-\frac{7579046612303}{990904320}\*\zeta_3\notag\\
&+\frac{14204709475283}{175177728000}\*\pi^4+\frac{2069200171}{10218700800}\*c_4
\,,\displaybreak[0]\\
  \bar{C}^{(3)}_{hA,4} ={}&\frac{4076693750761425008915897}{216048997032110161920000}-\frac{6057352065467958268127}{21773645455491072000}\*\zeta_3\notag\\
&+\frac{603806730856391}{143470559232000}\*\pi^4-\frac{5079437872417}{298896998400}\*c_4+\frac{25408}{693}\*\zeta_5
\,,\displaybreak[0]\\
  \bar{C}^{(3)}_{hF,4} ={}&-\frac{1238453334120405110098249}{223653206037381120000}-\frac{2700979154985278923783}{73630685115187200}\*\zeta_3\notag\\
&+\frac{284617888774393}{398529331200}\*\pi^4-\frac{2391601045007}{948879360}\*c_4
\,,\displaybreak[0]\\
  \bar{C}^{(3)}_{AA,4} ={}&\frac{1890270528083071670983183687}{1213920785999816294400000}-\frac{14356366246539791627425411}{1998223516049080320000}\*\zeta_3\notag\\
&+\frac{1401765535050243193001}{7785286426165248000}\*\pi^4-\frac{68889910850230336793}{64877386884710400}\*c_4\notag\\
&-\frac{14659863890116}{29462808375}\*\pi^4\*\log(2)+\frac{5593363801118}{93532725}\*\zeta_5+\frac{4512791972672}{29462808375}\*c_5\,,\displaybreak[0]\\
  \bar{C}^{(3)}_{AF,4} ={}&-\frac{7529404475657498198617187659}{1416240916999785676800000}+\frac{1775983388579435017337141}{25618250205757440000}\*\zeta_3\notag\\
&-\frac{22060794221193376773581}{19463216065413120000}\*\pi^4+\frac{150853987410708328753}{32438693442355200}\*c_4\notag\\
&+\frac{20846843102608}{9820936125}\*\pi^4\*\log(2)-\frac{169094614526984}{654729075}\*\zeta_5-\frac{5927657182976}{9820936125}\*c_5
\,,\displaybreak[0]\\
  \bar{C}^{(3)}_{FF,4} ={}&\frac{359753866350386757872033}{10626057300418560000}-\frac{1068896561883385053954563}{10407414146088960000}\*\zeta_3\notag\\
&+\frac{100109862435687089717}{76028187755520000}\*\pi^4-\frac{2049101914011949799}{405483668029440}\*c_4\notag\\
&-\frac{9491657579312}{4208972625}\*\pi^4\*\log(2)+\frac{15604738152184}{59520825}\*\zeta_5+\frac{17514775207168}{29462808375}\*c_5\,.
\end{align}

\bibliography{biblio}{}
\bibliographystyle{JHEP_modified}

\end{document}